\documentclass[english]{article}
\usepackage{amsmath}
\usepackage{amssymb}
\usepackage{cite}
\makeatletter
\usepackage{hyperref}

\usepackage{babel}

\begin{document}
\begin{titlepage}
\thispagestyle{empty}

\bigskip

\begin{center}
\noindent{\Large \textbf
{Null Second Order Corrections to Casimir Energy in Weak Gravitational Field}}\\

\vspace{0,5cm}

\noindent{A. P. C. M. Lima${}^{a}$\footnote{e-mail: augustopcml@gmail.com}, G. Alencar${}^{a}$, C. R. Muniz${}^b$  and R.R. Landim${}^{a}$}

\vspace{0,5cm}

{\it ${}^a$Departamento de F\'{\i}sica, Universidade Federal do Cear\'{a}-
Caixa Postal 6030, Campus do Pici, 60455-760, Fortaleza, Cear\'{a}, Brazil.

\vspace{0.2cm}
 }
 {\it ${}^b$Universidade Estadual do Cear\'a, Faculdade de Educa\c c\~ao, Ci\^encias e Letras de Iguatu,Av. D\'ario Rab\^elo, 63502-253 Iguatu, Cear\'a, Brazil}

\end{center}

\begin{abstract}
The discussion of vacuum energy is currently a subject of great theoretical importance, specially concerning the cosmological constant problem in General Relativity. From Quantum Field Theory, it is stated that vacuum states subject to boundary conditions may generate tensions on these boundaries related to a measurable non-zero renormalized vacuum energy: the Casimir Effect. As such, investigating how these vacuum states and energy behave in curved backgrounds is just natural and might provide important results in the near future. In this paper we revisit a model of the Casimir Effect in weak gravitational field background, which has been proposed and further generalized in the literature. A trick originally used to simplify calculations is shown to lead to a wrong value for the energy shift, and by performing explicit mode expansion we arrive at an unexpected result: null gravitational correction even at order $(M/R)^2$, in opposition to earlier results.
\end{abstract}
\end{titlepage}

\section{Introduction}
\label{sec:intro}

One rather odd aspect of Quantum Field Theory lies in the connection between quantum vacuum and the Casimir Effect \cite{casimir1948}: fluctuations of a confined field can generate measurable interactions even at vacuum states. Mathematically, this comes from the renormalization of the vacuum expected value of the stress-energy-momentum tensor in the presence of boundaries(or analogously topological identification conditions)\cite{Visser:2016ddm}. This quite general character makes it possible to extend the original model of electromagnetic fluctuations trapped between conducting surfaces for various fields, boundary conditions, surface geometries and spacetime topologies.

The first convincing experimental validations came about in 1997 \cite{lamoreaux} and by \cite{mohideen} in 1998. Although the model might seen theoretically simple, a number of practical dificulties arise when measuring casimir forces, such as deformation of the surfaces, corrections from finite conductivity, temperature, material boundary rugosity, among others. A few recent experiments are listed in references \cite{Chang2011,Chang:2012fh, Banishev:2012zz, Banishev:2012fi, Banishev:2012bh}. Unfortunately, as with many such measures expected at very high energy scales, present technology is very far from detecting tiny corrections such as those from extended standard-model theories and gravitational effects in the Casimir effect, though, it is possible to use current experimental data to impose bounds on parameters of the respective theories(more on this in the next paragraph).

For both theoretical and phenomenolgical interests, the study of Casimir Effect in non-trivial space-time topologies has gotten attention in the past decades. For being intrinsically related to quantum vacuum fluctuations (some authors say however that the effect is not actually related to zero point fluctuations, see for example \cite{Nikolic:2016kkp}), the investigation of the Casimir effect in curved background might be of cosmological value(look \cite{Martin:2012bt} for an general view at the cosmological problem). The subject of casimir interaction in curved backgrounds has already been extensively worked on, as can be exemplified from \cite{Bimonte:2008zva, dowker, Ford:1976fn,Altaie:1978dx,Baym:1993zx, Karim:2000ja, Setare:2001nx,Elizalde:2002dd,Aliev:1996va, Elizalde:2006iu, BezerradeMello:2006du, Elizalde:2009nt, NouriZonoz:2009zq, Nazari:2011hp, Saharian:2010ju, Bezerra:2011zz, Bezerra:2011nc, Milton:2011kz,BezerradeMello:2011nv, Sorge:2018zfd, Sorge, Nazari, stabile, Bezerra,Blasone:2018nfy,Buoninfante:2018bkc,Zhang,Bezerra:2016qof,Klimchitskaya:2014psa,Klimchitskaya:2015kxa, Sorge:2014uma, Muniz:2013uva, Quach:2015qwa}. As recent and potentially important phenomenological applications, we can highlight a few examples such as the bounds found on a Lorentz violation parameter \cite{Blasone:2018nfy}(which is lower than previously found ones), on one from Kehagias-Sfetsos solution for Horava-Lifshitz gravity\cite{Muniz:2013uva},  and a proposed test for the Heisenberg-Coulomb theory based on gravitonic casimir effect\cite{Quach:2015qwa}.

In this work we revisit the model of \cite{Sorge} where an expression for Casimir energy for a massless real scalar field in a weak static gravitational field is derived. Through a perturbative method, first order(in $M/R$) corrections are shown to be zero and a second order energy shift is found. Works from other authors generalizing the spacetime geometry are also exemplified \cite{Bezerra,Nazari,stabile,Buoninfante:2018bkc,Blasone:2018nfy}. Although the original approach is simple and elegant, in circumventing more lengthy calculations, an equivocated simplification is applied to obtain second order corrections in \cite{Sorge}, leading to a wrong value for the casimir energy. In this paper we rederive the results by performing explicit expansion of the mode solutions and carrying out the calculations directly, we show that the gravitational correction in the mean casimir energy density obtained by the model is in fact zero even at second order. The result is also extended for a more general spacetime equivalent to the one proposed at \cite{Nazari}. As a number of works followed from Sorge's paper, the present authors hope that the results presented here might be helpful in revising earlier ones, including those that might be of phenomenological importance like \cite{Blasone:2018nfy} as well as fueling pertinent questions to the general study of Casimir effect and quantum field theory on curved spacetime.

\section{Casimir Effect in a Weak Gravitational field-Brief review}\label{two}

In this section we present a review of the fundamentals for Casimir effect in curved space-time and a few results from the literature, we follow heavily \cite{Sorge}, which is the main reference of this paper.

\subsection{Real massless scalar field in curved background}
Consider a real massless scalar field is subjected to Dirichlet boundary conditions on rectangular parallel plates with a background static weak gravitational field. Measurements are taken by a static observer with four-velocity:
\begin{equation}\label{obs}
u^\mu=g_{00}^{-1/2}\delta_0^\mu.
\end{equation}
The action of the scalar field in curved background is given by:
\begin{equation}
S=\int d^4 x \sqrt{g}[\frac{1}{2}g^{\mu\nu}\partial_\mu \phi\partial_\nu\phi+\frac{1}{2}\epsilon R \phi^2],
\end{equation}
so that the field equation in curved space-time is:
\begin{equation}\label{KG}
\frac{1}{\sqrt{g}}\partial_\mu[\sqrt{g}g^{\mu\nu}\partial_\nu\phi]+\epsilon R \phi=0.
\end{equation}
To find a simple expression for the energy density, first we work out orthonormal mode solutions in the sense of the scalar product:
\begin{equation}\label{A}
\langle \phi_n , \phi_m \rangle = \int_\Sigma \sqrt{g_\Sigma}n^\mu [\phi_n^*\partial_\mu\phi_m-(\partial_\mu\phi_n^*)\phi_m]d \Sigma,
\end{equation}
where $\Sigma$ stands for the boundaries, and
\begin{equation}\label{normal}
\langle \phi(n,\vec{k}_n), \phi(m,\vec{k}_m)\rangle=\delta(\vec{k}_n-\vec{k}_m)\delta_{nm}.
\end{equation}
Here the $\vec{k}$ terms refers to the transverse wave numbers and the $n,m$ are discrete modes generated by the boundary condition impositions on the field. In terms of these orthonormal modes the mean vacuum energy can be expressed as:
\begin{equation}\label{energy}
\bar{\epsilon}=\frac{1}{V_p}\sum_n\int d^2k \sqrt{g_\Sigma}(g_{00})^{-1}T_{00}[\psi_n^*,\psi_n],
\end{equation}
where $V_p$ is the proper volume of the cavity and $T_{00}[\psi_n^*,\psi_n]$ is defined after the energy-momentum component:
\begin{equation}\label{tens}
T_{00}[\psi_n^*, \psi_n]=\partial_0\psi_n^* \partial_0\psi_n-g_{00}g^{\mu\nu}\partial_\mu\psi_n^*\partial_\nu\psi_n.
\end{equation}

\subsection{Casimir energy for the the flat space-time case}

In the case $g_{\mu\nu}=\eta_{\mu\nu}$, the field is governed by the usual Klein-Gordon equation. The boundaries are planes of coordinate separation L such that $\phi(z=0)=\phi(z=L)=0$, orthonormal solutions are given by:
\begin{equation}
\phi(x)=\frac{1}{2\pi\sqrt{\omega_{nF} L}}\sin\left(\frac{n\pi z}{L}\right)\exp[i(\omega_{nF} t-k_\perp x_\perp)],
\end{equation}
where:
\begin{equation} 
\omega_{nF}=\sqrt{k_\perp^2+(n\pi/L)^2}
\end{equation}
Is the mode frequency($\mathit{F}$ standing for flat). Applying these modes on equation (\ref{energy}) results in:
\begin{equation}\label{flatt}
\bar{\epsilon}=\frac{1}{8\pi^2 L}\sum_n\int d^2k_\perp \omega_{n}F.
\end{equation}
This integral diverges as expected. Using Schwinger's proper time representation and zeta function regularization we obtain the renormalized value for the Casimir energy in minkowsi space-time:
\begin{equation}\label{flat}
\bar{\epsilon}_{cas}= -\frac{\pi^2}{1440L^4}.
\end{equation}

\subsection{Sorge's result}\label{sorges}
In reference \cite{Sorge}, from F. Sorge, the Casimir energy is calculated  perturbatively for the following weak gravitational field space-time metric:
\begin{equation}\label{weak}
ds^2=-(1+2\Phi)dt^2+(1-2\Phi)dl^2,
\end{equation}
for rectangular Dirichlet boundaries of coordinate surface $S$ and coordinate separation $L$. The factor $\Phi$ is expanded inside the cavity (to order $(M/R)^2$) as:
\begin{equation}
\Phi= -M/r \simeq -M/R+Mz/R^2= \Phi_0+\gamma z.
\end{equation}
Introducing a rectangular coordinate system inside the cavity. The center of the inner plate has spatial coordinates $(0,0,0)$ while the center of the outer one has $(0,0,L)$. We consider $L\ll\sqrt{S},M$, so that the plates are finite but we can still approximate transverse modes by plane waves.

The line element then looks like
\begin{equation}\label{line}
ds^2=-(1+2\Phi_0+2\gamma z)dt^2+(1-2\Phi_0-2\gamma z)dl^2,
\end{equation}
with a coordinate change
\begin{align}
dt \rightarrow (1+2\Phi_0)^{-1}dt\\
\vec{dx}\rightarrow (1-2\Phi_0)^{-1}\vec{dx},
\end{align}
and we arrive at
\begin{equation}\label{metric}
ds^2=-(1+2\gamma z)dt^2+(1-2\gamma z)dl^2.
\end{equation}
The line element now is the same as (\ref{line}) with $\phi_0=0$. This is to tell us that up to order $[M/R]^2$ the parameter of interest is $\gamma$, that of the spatially varying field. It is shown in \cite{Zhang} that as long as there are no cross terms in the metric, the Casimir energy for a general constant perturbation measured by observer (\ref{obs}) is unchanged, this is physically expected, since those space-times are equivalent to the Minkowski one trough a coordinate reescaling.

The case $\gamma=0$, or first order aproximmation is calculated explicitly an shown to be null in \cite{Sorge}. Next, for second order correction metric(\ref{metric}) is considered. For this spacetime the field equation becomes
\begin{equation}\label{KGE}
-(1-4\gamma z)\partial_t^2\phi+\nabla^2\phi=0,
\end{equation}
where $\nabla^2\equiv \delta^{ij}\partial_i\partial_j$. With mode solutions of the form
\begin{equation}\label{modess}
\phi_{n,k}=\chi_n(z)e^{i\omega_{n} t -i k_\perp x_\perp}.
\end{equation}
The assymptotic solutions are
\begin{equation}\label{mod}
\chi(u)=A_n u^{-1/4}\sin\left(\frac{2}{3}u^{3/2}+\varphi\right),
\end{equation}
where:
\begin{align}\label{udz}
u(z)=&-(z-b/a)a^{1/3}\nonumber\\
a=&4\gamma\omega_n^2\\
b=&\omega_n^2-k_\perp^2\nonumber.
\end{align}
And:
\begin{equation}
\omega_n=(1+\gamma L)\omega_{nF}
\end{equation}
Are the frequencies. Instead of working the modes explicitly the author considers an expansion in terms of the flat case solution(\ref{flat})
\begin{equation}
\phi_n=\phi_{n}^{(0)}+\delta\phi_n.
\end{equation}
Then, the energy-momentum tensor and Casimir energy are expanded as:
\begin{equation}\label{deco}
T_{00}[\phi_n^*,\phi_n]=T_{00}[\phi_{n}^{(0)*},\phi_{n}^{(0)}]+\{T_{00}[\delta\phi_n^*,\phi_{n}^{(0)}]+c.c\},
\end{equation}
and:
\begin{equation}
\bar{\epsilon}=\bar{\epsilon}^{(0)}+\delta\bar{\epsilon}.
\end{equation}
By substituting the first term of the RHS of (\ref{deco}) on (\ref{energy}) and renormalizing we obtain:
\begin{equation}
\bar{\epsilon}^{(0)}_{(ren)}=-(1-2\gamma L_p)\frac{\pi^2}{1440L_p^4},
\end{equation}
and the second term, in analogy to the flat space-time expression(\ref{flatt}), is to be calculated by integrating the frequency change of the modes(other geometrical terms can be neglected since $\delta\phi$ itself is of order $\gamma$):
\begin{equation}
\delta\bar{\epsilon}=\frac{1}{8\pi^2 L}\sum_n\int d^2k_\perp \delta\omega_n=\gamma L_p\left(-\frac{\pi^2}{1440L_p^4}\right),
\end{equation}
arriving at
\begin{equation}
\bar{\epsilon}_{casimir}=-(1-\gamma L_p)\frac{\pi^2}{1440L_p^4},
\end{equation}
which is the original result presented by the author.

\subsection{Generalizations}
In this subsection we briefly review a few generalizations proposed in the literature to the above result, calculations will not be shown explicitly, since accurate derivations of the Casimir energy will be presented in the next section.
 In \cite{Bezerra} the authors consider the far field limit in the case of a rotating source, the metric used is
\begin{equation}
ds^2=(1+2\phi)dt^2-(1-2\phi)dl^2-4adtd\varphi,
\end{equation}
where $b=1-2a\Omega$, which is related to the rotation parameters. Through coordinate change it is cast into (first order terms are eliminated similarly to Sorge's case)
\begin{equation}
ds^2=(1+2b\gamma z)dt^2-(1-2\gamma z)dl^2.
\end{equation}
The procedure from last subsection is followed tightly, the mean casimir energy density is fond to be
\begin{equation}
\bar\epsilon_{cas}=-[1-\gamma L_p(1+3a\Omega)]\frac{\pi^2}{1440L_p^4}.
\end{equation}
Thermal corrections are also calculated but we will omit them for conciseness.

In \cite{Nazari} the space-time metric is proposed
\begin{equation}
ds^2=(1+2\gamma_0+2\lambda_0 z)dt^2-(1+2\gamma_1+2\lambda_1 z)dl^2,
\end{equation}
which is then simplified to Fermi-Coordinates assuming the form
\begin{equation}
ds^2=(1+2\lambda_0 z')dt^2-dl'^2.
\end{equation}
Notice that the parameters from the spacial part of the metric are eliminated from the metric, though they will be present in the boundary conditions. A general analysis is provided for scalar and vector fields with both Neumann and Dirichlet conditions, for our case (scalar field and Dirichlet boundaries) the result presented for the mean casimir energy is
\begin{equation}
\bar\epsilon_{cas}=-(1+\gamma_0+\lambda_0\frac{L_p}{2})\frac{\pi^2}{1440L^4}.
\end{equation}
Note that only the denominator is expressed in terms of the proper length.

In \cite{stabile} the background space-time is that of Extended Theories of Gravity (see the reference for more detailed discussion and explanation on the parameters used). The metric is put in a familiar form:
\begin{equation}
ds^2=(1+2\Phi_0+2\Lambda z)dt^2-(1-2\Psi_0-2\Sigma z)dl^2.
\end{equation}
The authors explain that in this theory there is a vaccum curvature scalar $\mathcal{R}\simeq \mathcal{R}_1+\mathcal{R}_2 z$, so geometric coupling is considered in equation (\ref{KG}). The result obtained is
\begin{equation}
\bar{\epsilon}_{cas}=-[1-3(\Phi_0+\Psi_0)-(2\Sigma-\Lambda)L_p]\frac{\pi^2}{1440L_p^4}+\frac{\epsilon\mathcal{R}_2}{192L_p}.
\end{equation}

The Casimir effect analysed in the recent paper \cite{Buoninfante:2018bkc} it is quite similar to another reference that approaches Extended Theories of Gravity \cite{stabile,Blasone:2018nfy}. We found to be difficult to reproduce their results because in the Casimir effect part the authors practically only exhibit both the solution in Sorge modes and the normalization constant, and then they already present the result of the energy without greater details.

Notice that the afore mentioned results do not coincide if the proper parameter equivalence is established, also first order corrections are found, what is not in concordance with the arguments for first order correction in \cite{Sorge}. In the next section we will present explicit calculations to properly review these results.

\section{Revisiting}\label{three}
Let us first review Sorge's trick for second order calculations. The same arguments which led to it could be used to compute the first order correction as well, let us see what it gives. For $\Phi=\Phi_0$ in (\ref{weak}), substituting mode (\ref{flat}) on (\ref{tens}) and integrating in (\ref{energy}) we get
\begin{equation}
\bar{\epsilon}^{0}=\frac{1}{V_p}\frac{1}{(2\pi)^2 2l}(1-3\Phi_0)\sum_n\int d^2k \omega_0.
\end{equation}
Renormalizing and expressing in terms of $L_P$ it becomes
\begin{equation}
\bar{\epsilon}^{0}_{ren}=-(1-4\Phi_0)\frac{\pi^2}{1440L_p^4}.
\end{equation}
The frequencies are $\omega\simeq (1+2\Phi_0)\omega_0$ so we should have
\begin{equation}
\delta\bar{\epsilon}=-2\Phi_0\frac{\pi^2}{1440l_p^4}.
\end{equation}
Summing both contributions gives a nonzero energy shift. The method goes wrong on the assumption that $\delta\bar{\epsilon}$ can be calculated by substituting $\omega_0$ for $\delta\omega$ in the flat case expression. We will show later that the correction factor is actually connected to a geometrical term reminiscing normalization condition(\ref{A}).

\subsection{Explicit calculation of second order corrections to the Casimir Energy}
We begin by casting (\ref{mod}) into a more familiar expression by rewriting this mode axial solution in terms of the flat space-time mode solution plus a perturbative term
\begin{equation}\label{assynt}
\chi(z)=\frac{1}{2\pi\sqrt{\omega_0 L}}\sin\left(\frac{n\pi z}{L}\right)+\gamma\chi^{(1)}(z)+O[M/R]^3.
\end{equation}
For that we need to find $\varphi$,$A_n$ and $\omega_n$. The frequencies are obtained by requiring periodicity of the solution ($\psi(0)=\psi(L)=0$), according to
\begin{equation}
\omega_n\simeq(1+\gamma L)\omega_0,
\end{equation}
where $\omega_0=[k^2+(n\pi/L)^2]^{1/2}$. Requiring that $\chi(z=0)=0$ immediately gives
\begin{equation}
\phi=-\frac{2}{3}u^{3/2}(0).
\end{equation}
For $A_n$ we first express (\ref{A}) taken on a hypersurface $t=0$ in terms of $\chi$
\begin{align}\label{normal}
&\langle \psi(n,k_1),\psi(n,k_2) \rangle=\nonumber\\&(\omega_n+\omega_m)\int_V d^3x (1-4\gamma z)\chi_n\chi_m e^{i(k_{1\perp}-k_{2\perp})x_\perp}=\nonumber\\
&\delta^2(k_{1\perp}-k_{2\perp})\delta_{nm},
\end{align}
which leads to
\begin{equation}
A_n=\left[(2\pi)^2 2\omega\int_V d^3x (1-4\gamma z))\Theta_n^2]\right]^{-1/2},
\end{equation}
where $\Theta_n=u^{-1/4}\sin\left(\frac{2}{3}u^{3/2}+\varphi\right)$. Plugging the result on the mode solution leads to the correction
\begin{align}
\chi^{(1)}(z)=[2n\pi\omega_0^2L^2  (L-z)z\cos(n\pi z/L)\nonumber\\
+L(2n^2\pi^2 z+ 2 k^2 L^2 z- k^2 L^3)\sin(n\pi z/L)]/4 L n^2 \pi^3\sqrt{\omega_0 L}.
\end{align}
With the full form of the mode solution (\ref{assynt}) it can checked that equation (\ref{KG}), as well as boundary and normalization conditions are satisfied to order $(M/R)^2$.

Now all that is left is to use the definition (\ref{energy}). For $\chi(z)$, equation (\ref{tens}) reads
\begin{equation}\label{EM}
T_{00}[\psi_n^*, \psi_n]=\frac{1}{2}\omega_n^2\chi_n^2+\frac{1}{2}(1+4\gamma z)[k_\perp^2\chi_n^2+(\partial_z\chi_n)^2].
\end{equation}
The unnormalized vacuum energy density is
\begin{equation}
\bar{\epsilon}=\frac{1}{V_p}\frac{S}{8\pi^2 L}(1+\gamma L/2)\sum_n\int d^2k_\perp\omega_0.
\end{equation}
The standard proper-time representation for the $k_\perp$ integration and Riemann zeta function regularization lead to the renormalized Casimir energy density
\begin{equation}\label{energy1}
\bar{\epsilon}_{cas}=-\frac{\pi^2}{1440L^4}(1+2\gamma L)=-\frac{\pi^2}{1440L_p^4}.
\end{equation}
So, in terms of the proper(measured by static observer) length of the cavity, the mean casimir energy density (and hence total energy) is just the same as the flat space-time case for a rigid cavity.

Either the expression (\ref{energy1}) tells us something important or not is tricky to say, since in generalized coordinates it becomes a lot more difficult to distinguish between actual physical effects and those simulated by choice of coordinates. The``measured" casimir energy density looking the same as that of flat space-time could be a effect of not considering covariant boundary conditions (remember that conditions are imposed on the coordinate surface and separation for the plates) for example. Putting those arguments aside, from a heuristic point of view it is interesting that the shift in the field solution is found in this calculation to be just enough to compensate for the geometrical factors.

Although the result is disappointing from the phenomenological point of view, gravitational corrections are still expected on thermal effects, non static spacetimes such as Kerr's\cite{Sorge:2014uma} as well as higher order corrections (sadly, even corrections of order $(M/R)^2$ are  far from what would be detectable with current technology).

\subsection{Generalization to spherically symmetrical space-time}

A general spherically symmetrical space-time line element looks like(in isotropic coordinates)
\begin{equation}\label{ST}
ds^2=-e^{2\phi_t(r)}dt^2+e^{-2\phi_s(r)}(dr^2+r^2d\Omega^2),
\end{equation}
where $\phi_s$ e $\phi_t$ are general functions of $r$ (not (r,t), since by Birkhoff's theorem any spherically symmetrical solution will also be static.). In the same fashion as the last section we can treat this problem (in weak field regime) using the line element
\begin{equation}\label{metric2}
ds^2=-(1+2\gamma_t z)dt^2+(1-2\gamma_s z)dl^2.
\end{equation}
The EoM for the scalar field is
\begin{equation}
-(1-2\gamma_t z)\partial_t^2\phi+(1+2\gamma_s z)\nabla^2\phi+(\gamma_t-\gamma_s)\partial_z\phi=0.
\end{equation}
With solutions of the form (\ref{modess}) we get
\begin{equation}
\partial_z^2\chi+2\gamma_-\partial_z\chi+(\omega^2-k_\perp^2)\chi-4\gamma_+\omega^2z\chi=0,
\end{equation}
where $\gamma_+=(\gamma_s+\gamma_t)/2$ and $\gamma_-=(\gamma_t-\gamma_s)/2$. The solutions are
\begin{equation}\label{modez}
\chi(z)=A_n(1-\gamma_- z)\Theta(u(z)),
\end{equation}
Where $\Theta(u)$ and frequency modes are defined in analogy to the last section with $\gamma\rightarrow\gamma_+$.

It will be instructive to keep a couple terms implicit in the calculation. Let (\ref{normal}) be rewritten as
\begin{align}
&\langle \phi(n,k_1),\phi(n,k_2) \rangle=\nonumber\\&(\omega_n+\omega_m)
\int_V d^3x (1-\beta z)\chi_n\chi_m e^{i(k_{1\perp}-k_{2\perp})x_\perp},
\end{align}
and (\ref{energy}) as
\begin{equation}\label{energy2}
\bar{\epsilon}=\frac{1}{V_p}\sum_n\int d^3x\int d^2k (1-\lambda z)T_{00}[\psi_n^*,\psi_n].
\end{equation}
Also, $\gamma$ is replaced by $\gamma_+$ in (\ref{EM}). Applying the solution (\ref{modez}) to the corresponding expressions leads to the renormalized energy in the form
\begin{equation}
\bar{\epsilon}_{ren}=-\frac{S\pi^2}{1440L^3}\frac{1}{V_p}\left[1+\frac{1}{2}(\beta-\lambda+2\gamma_+)L\right]=-\frac{\pi^2}{1440L_p^4}.
\end{equation}
In this form it is easier to see where each term comes from. One can check that inserting the Minkowski solution in (\ref{energy2}) leads to the $\gamma_+$ and $\lambda$ term while the correction associated exclusively to the shift in the modes is
\begin{equation}
\delta\bar{\epsilon}_{ren}=\beta L\epsilon_0,
\end{equation}
which for the original case was $\delta\bar{\epsilon}=2\gamma\epsilon_0$. That leads to the zero correction of our calculation. So the multiplicative factor involved is not directly associated to the frequency correction as assumed in \cite{Sorge}, but rather to the spatial metric from the normalization integral. So,also for this case no gravitational correction is expected. Of course, this is for zero temperature, which is not the case in real world experiments, for example, gravity terms are still expected in thermal corrections.

\section{Conclusions}\label{conc}

In this paper we have revisited the problem of a Casimir apparatus of parallel plates in the weak gravitational background described by (\ref{metric}). We followed the model from \cite{Sorge}, now deriving the second order mode solution explicitly, finding a null correction for the mean Casimir energy density, in variance with the former author and, consequently, with all the works based in it, including the more recent ones which take into account extended theories of gravity \cite{Blasone:2018nfy,Buoninfante:2018bkc}. The method originally used was discussed and shown to fail when applied to the first order correction. We generalize this result to a static spherically symmetrical space-time(\ref{metric2}). The resulting corrections are also zero, in variance with \cite{Nazari, stabile}.

From a phenomenological point of view, the results obtained here are not as interesting as the previous ones since there is no actual energy shift, although corrections may appear when considering thermal effects, finiteness and border effects of the boundaries, as well as higher corrections. An useful analysis that should to be made in the sense here discussed is with respect to the stationary (Kerr-like) space-times, which according to the literature already presents first order gravitational corrections in the Casimir energy density\cite{Sorge:2014uma}, this is in accordance with \cite{Zhang}, since those effects can show up at non-static spacetimes even if metric coefficients are constant.
 
The odd result of absence of gravitational correction for the cases considered might be interesting though, from a theoretical point of view. It is curious  that the action of the curved background up to second order is to exactly cast coordinate separation into proper distance as shown in equation (\ref{energy1}). It also raises the question that, should the static gravitational field(\ref{metric2}) generate no correction to the Casimir Energy up to second order, then, on which order(or if) should those effects actually appear. 

Although the obtained results appear to partially frustrate near future attempts of measuring gravitational changes in the Casimir effect, the interpretation of these results is to be tackled cautiously, and the question as to how vacuum energy fluctuations behave in curved space-times for simpler cases such as weak-field approximation might provide useful hints for the understanding of wider pictures. As such, the present authors hope to aid further discussions on the subject.

\section*{Acknowledgments}
The authors would like to thanks Alexandra Elbakyan and sci-hub, for removing all barriers in the way of science. 

We acknowledge the financial support  by the Conselho Nacional de Desenvolvimento Cient\'ifico  e Tecnol\'ogico(CNPq) and Funda\c{c}\~ao Cearense de Apoio ao Desenvolvimento Cient\'ifico e Tecnol\'ogico(FUNCAP) through PRONEM PNE0112-00085.01.00/16.

%\bibliographystyle{unsrt}
%\bibliography{Article}

\end{document}